\documentclass[twocolumn,superscriptaddress]{revtex4}
\usepackage{graphicx}
\usepackage{epstopdf}
\usepackage{dcolumn}
\usepackage{bm}
\usepackage{amsthm}
\usepackage{amsfonts}
\usepackage{color} 
\usepackage[usenames,dvipsnames]{xcolor}
\usepackage{amscd}
\usepackage{amsmath}
\usepackage{enumerate}
\usepackage{bbm}
\usepackage[colorlinks=true,citecolor=blue,urlcolor=black]{hyperref}
\usepackage{comment}
\usepackage{float}
\usepackage{mathtools}
\usepackage{amsthm}
\usepackage{subfigure}
\usepackage{dsfont}

\newcommand{\be}{\begin{equation}}
\newcommand{\ee}{\end{equation}}

\newcommand{\sket}[1]{{\ensuremath{\lvert#1\rangle}}}
\newcommand{\lket}[1]{{\ensuremath{\left\lvert#1\right\rangle}}}
\newcommand{\ket}[1]{\if@display\lket{#1}\else\sket{#1}\fi}

\newcommand{\sbra}[1]{{\ensuremath{\langle#1\rvert}}}
\newcommand{\lbra}[1]{{\ensuremath{\left\langle#1\right\rvert}}}
\newcommand{\bra}[1]{\if@display\lbra{#1}\else\sbra{#1}\fi}

\newcommand{\sbraket}[2]{{\ensuremath{\langle#1\rvert#2\rangle}}}
\newcommand{\lbraket}[2]{{\ensuremath{\left\langle#1\!\left\rvert\vphantom{#1}#2\right.\!\right\rangle}}}
\newcommand{\braket}[2]{\if@display\lbraket{#1}{#2}\else\sbraket{#1}{#2}\fi}

\newcommand{\sketbra}[2]{{\ensuremath{\lvert #1\rangle\!\langle #2\rvert}}}
\newcommand{\lketbra}[2]{{\ensuremath{\left\lvert #1\right\rangle\!\!\left\langle #2\right\rvert}}}
\newcommand{\ketbra}[2]{\if@display\lketbra{#1}{#2}\else\sketbra{#1}{#2}\fi}

\newcommand{\coloneqq}{:=}

\theoremstyle{nonumberplain} 
\newtheorem*{claim*}{Claim}

\newtheorem*{cor*}{Corollary}

\theoremstyle{definition}

\begin{document}

\title{Beating direct transmission bounds for quantum key
distribution with a multiple quantum memory station}
\begin{abstract}
Overcoming repeaterless bounds for the secret key rate capacity of quantum key distribution protocols is still a challenge with current technology. D. Luong \textit{et} \textit{al.} [Applied Physics B 122, 96 (2016)] proposed a protocol to beat a repeaterless bound using one pair of quantum memories. However, the required experimental parameters for the memories are quite demanding. We extend the protocol with multiple pairs of memories, operated in a parallel manner to relax these conditions. We quantify the amount of relaxation in terms of the most crucial memory parameters, given the number of applied memory pairs. In the case of high-loss channels we found that adding only a few pair of memories can make the crossover possible.
\end{abstract}
\author{R\'obert Tr\'enyi}
\affiliation{Escuela de Ingenier\'ia de Telecomunicaci\'on, Dept. of Signal Theory and Communications, University of Vigo, E-36310 Vigo, Spain}
\author{Norbert L\"utkenhaus}
\affiliation{Institute for Quantum Computing and Department of Physics and Astronomy,
University of Waterloo, Waterloo, Ontario, Canada N2L 3G1}

\maketitle

\section{Introduction}
Quantum Key Distribution (QKD) protocols~\cite{GisinReview,NorbertReview,MarcosReview} can establish information-theoretically secure secret keys between two parties. However, we notice that in optical implementations, the achievable key rate scales roughly with the single photon transmittance $\eta$, which leads to an exponential drop of the key rate with distance in optical fibers. This is not just a lack of good protocols, but it has indeed been proven that the secret key rate capacity of direct transmission QKD protocols is upper bounded by this scaling~\cite{Pir0,TGW,AnotherPaperOnBounds,PLOB,Pir1}. In this work we will focus on the so-called Pirandola-Laurenza-Ottaviani-Banchi (PLOB)~\cite{PLOB} bound, which is given by the formula  $-\log_2\left[1-\eta\right]$ and scales as $\eta$ for small transmission $\eta$ (\textit{i.e.}, long distances).

It is clear that intermediate nodes have to be added to go beyond the PLOB bound. One recent successful approach to beat the PLOB bound has been the development of Twin-Field QKD~\cite{LucamariniTfQkd}, which uses one untrusted intermediate station to achieve a key rate scaling of $\sqrt{\eta}$. Security proofs~\cite{MarcosTf,CuiTF,JieTF} and proof-of-principle experiments~\cite{ProofOfPrincipleTF,ToshibaExpTF,PanTF,WangTFExp} have already appeared in the literature. 

There have been many proposals constructed in a way that signals from the parties do not have to arrive at the same time to the middle station, one possibility is to utilize quantum memory assisted protocols containing one middle node~\cite{Luong,abruzzo, panayi, razavi}, which can also offer a square root improvement in the secret key rate capacity since independent successful arrival of the signals to middle station suffices for key generation. Also, a scheme without the need for quantum memories has been introduced in~\cite{koji} offering the same square root improvement using optical switches and the idea of multiplexing. 

In general, full-scale quantum repeaters~\cite{BriegelRepeater,DuanRepeater,HybridRepeater,KokRepeater} could be the final solution for long distance quantum communication, which, in principle, can provide better than a square root improvement depending on the number of nodes applied.

The main goal of this paper is to explore the parameter regime where a cross-over between a single-node quantum repeater and the PLOB bound on direct transmission can be realized. This approach will serve as a benchmark that pushes technology towards a fully scalable multi-node quantum repeater. Our approach is to extend the protocol~\cite{Luong} with the idea of using more quantum memories at the middle station and operating them in a parallel manner (using a multiplexing technique), which was also applied in~\cite{koji,ChristophSimon, razaviMultiple}. We mention that in~\cite{ChristophSimon, razaviMultiple} this idea has been used in a multi-node repeater setting. Here we evaluate the performance of this extension of~\cite{Luong} focusing on the most crucial imperfections of a quantum memory assisted system, \textit{i.e.}, dephasing~\cite{dephasingproblem}, non-unit efficiency of entanglement preparation, photon-fiber coupling and wavelength conversion. We quantify how an increasing number of quantum memories reduces the challenge on the required memory parameters for overcoming the PLOB bound compared to the one memory pair case~\cite{Luong}. We will also compare the performance of this extension to the upper bound for the secret key rate capacity of any QKD protocol containing one middle node, placed halfway between the parties~\cite{pirandolaGeneralBound}. 

The paper is organized as follows. In Sec.~\ref{Sec:Model} we describe the investigated protocol together with the mathematical model of the components used for its implementation. Then, in Sec.~\ref{methods} the secret key rate of the protocol is addressed. The main results of the paper are presented in Sec.~\ref{Sec:results}. Finally, we conclude the paper in Sec.~\ref{Sec:Conclusion}. The paper also contains an Appendix, in which the derivation of the secret key rate formula is given.

\section{Model of the scheme}\label{Sec:Model}
 
\begin{figure}[H]
  \centering
  \includegraphics[scale=0.35]{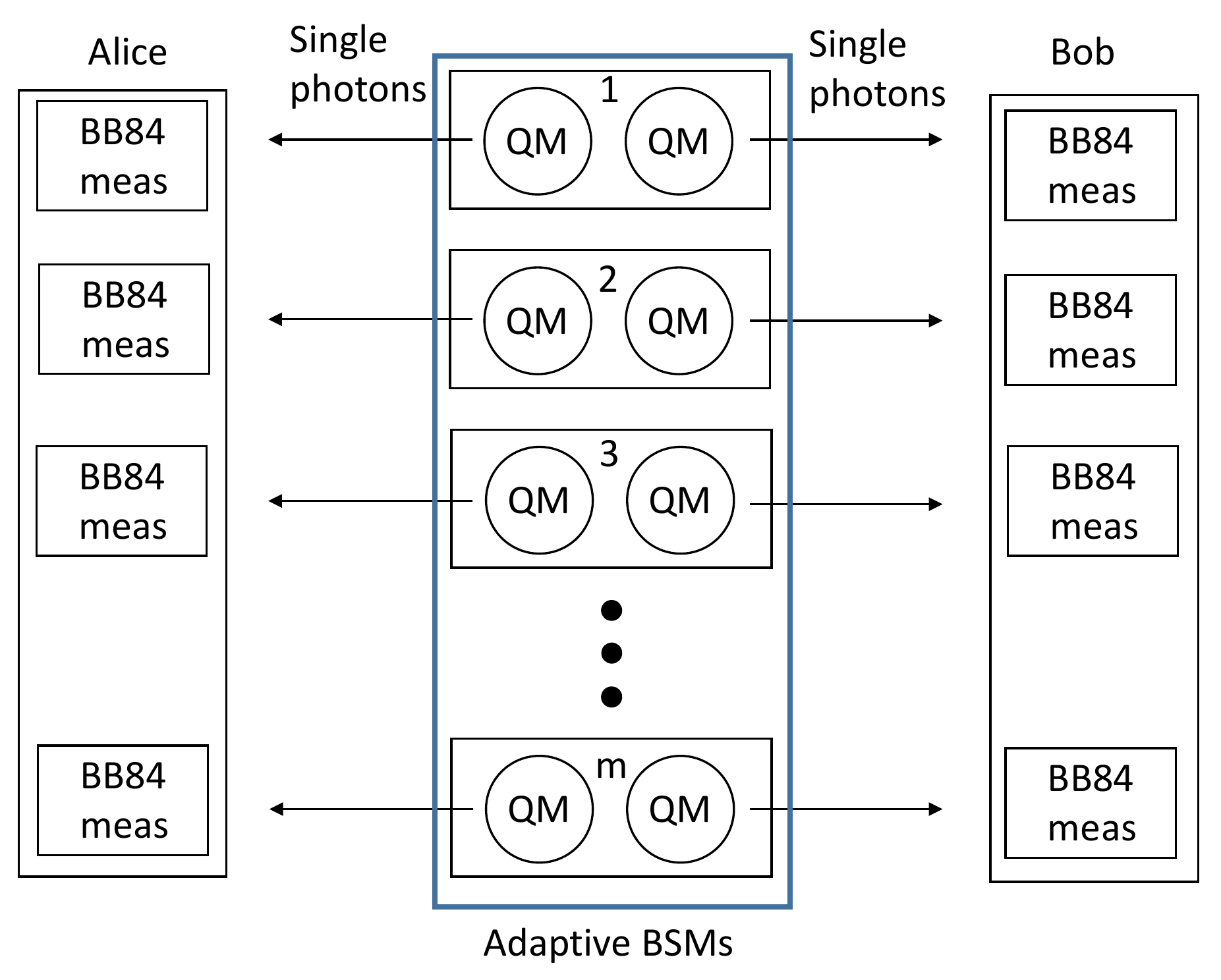}
  \caption{Schematic layout of the investigated protocol having altogether $m$ memory modules, forming the middle station, operating in a parallelized manner. Each memory module contains two quantum memories (QM), the one on the left (right) emits signals towards Alice (Bob). The central station is located halfway between Alice are Bob, who are separated at a distance $L$. The parties are connected to the central station via lossy bosonic channels. Bell state measurements (BSM) can be performed between the QMs on the right and the left. Note, that QMs are paired for the BSMs via a switch so it is possible that QMs located within different memory modules are paired.}
  \label{fig:protocol}
\end{figure}

The schematic layout of the protocol investigated here is depicted in Fig.~\ref{fig:protocol}. This protocol can be considered as an extension of the protocol proposed in~\cite{Luong} to operate $m>1$ memory modules at the same time in a parallel manner. This parallelization idea was also used in~\cite{koji}. 

Each memory module, labeled by different numbers from 1 to $m$ in Fig.~\ref{fig:protocol}, consists of two quantum memories (QM). A QM is capable of emitting single photons, which are entangled to the QM. Moreover, as in~\cite{Luong}, we assume, that Bell state measurements (BSM) can be performed between the QMs. It is important to note, that since we have more memories than in the original proposal we require a switching mechanism, that allows pairing between QMs that can be located in different memory modules, labeled by different numbers. We remark that this switching mechanism has been employed before, for instance in the  protocol investigated in~\cite{dephasing}, which contains a chain of quantum memory stations. It is also necessary for the fully optical, adaptive MDI-QKD protocol proposed in~\cite{koji}.  

\subsection{Protocol steps}
The protocol is composed of rounds. We define a round to be the process when all the $m$ memory modules emit single photons simultaneously towards one party. The steps of the protocol:

\begin{enumerate}
\item For each individual signal, coming from different QMs, Alice decides to measure it either in the $Z$ or $X$-basis with probabilities $p_{\rm{Z}}$ and $p_{\rm{X}}=1-p_{\rm{Z}}$. It is assumed that Alice chooses the $Z$ basis much more often than the $X$-basis (i.e., $p_{\rm{Z}}\gg p_{\rm{X}}$), as in~\cite{effqkd}. Alice repeats the rounds successively until she detects at least one single photon on her side. The QMs corresponding to the successful detections are considered to be loaded. We note that a similar step is also used in the protocol presented in~\cite{ChristophSimon,dephasing}.
\item After Alice succeeded, Bob does the same process with the QMs on his side.
\item The loaded memories from the two sides are paired via a switching technique and Bell state measurements (BSM) are performed between them. Note that due to the switching mechanism applied, the paired QMs can be located in different memory modules, labeled by different numbers in Fig.~\ref{fig:protocol}. 
\item Then the results of the BSMs are announced and the raw key is obtained as in the original MDI-QKD~\cite{mdi}.
\item The leftover loaded memories, for which it was not possible to find a pair, are erased and steps 1-4 are repeated until sufficient amount of raw key is obtained. 
\end{enumerate}

We point out that it might be possible to improve the performance of the protocol if the unpaired memories from step 5 are not thrown away, however, these memories will continue dephasing until the other party loads some of her/his QMs and pairings can start. Therefore, we do not expect a significant performance boost by keeping these QMs. We note that by the key rate of the protocol we actually mean the key rate per mode in order to be able to make a fair comparison to the PLOB bound. Therefore, the fact that BB84 measurements are used, brings in a factor of $1/2$ in the secret key rate since two optical modes are required for BB84.

\subsection{Modeling of components}\label{params}
We reuse the description of all the components used in~\cite{Luong}. This is reasonable since when we look at a particular pairing, that is essentially the same process which was considered in the original proposal. The only additional component that we apply here is a device (e.g. a switch), that is capable of pairing~\cite{poster} the successfully loaded quantum memories (even if they are located in different memory modules) to perform the BSMs between them. For simplicity, we assume this device to be perfect, that is, it is lossless and the pairing can be done instantly. To have all the parameters at hand describing the imperfections of our implementation, we briefly repeat the description of the components in~\cite{Luong}.

\subsubsection{Quantum memories}
The QM can generates the photon-memory entangled state $\ket{\phi^{+}}$ with success probability $\eta_p$. Each trial of generating this maximally entangled state needs a preparation time of $T_p$.

The memory-channel coupling efficiency is denoted by $\eta_c$, which includes the probability that the emitted photon will enter the channel and also the probability of a successful wavelength conversion, if it is necessary.

The dephasing time constant of the QM is $T_2$. Basically, $T_2$ characterises how fast the state stored by the QM deviates from the original, maximally entangled state. The dephasing is modelled by the following map~\cite{dephasing}, that gives the quantum state of the QM after $t$ time has passed given its initial state $\rho$:
\begin{equation}\label{dephasing}
\Gamma_t(\rho)=(1-\lambda_{dp}(t))\rho+\lambda_{dp}(t)Z\rho Z,
\end{equation}
where $Z$ is the Pauli $Z$ operator and
\begin{equation}\label{dephasingLambda}
\lambda_{dp}(t)=\frac{1-e^{-t/T_2}}{2}.
\end{equation}
\subsubsection{Optical channels}
Alice and Bob are both connected to the central memory station via optical fibers of length $L/2$, which means that the distance between them is $L$. The speed of light in the optical fiber is denoted by $c$. The transmittance of a fiber of length $l$ is given by
\begin{equation}\label{transmittivity}
\eta_{ch}=\exp\left({-l/L_{att}}\right),
\end{equation}
where $L_{att}$ is the attenuation length of the fiber.

Due to the misalignment in the channel between Alice and the middle station, the initial photon-memory state $\ket{\phi^{+}}$ is transformed into the following state after passing through the channel: 
\begin{equation}\label{misalignment}
(1-e_{mA})\ket{\phi^{+}}\bra{\phi^{+}}+e_{mA}\ket{\psi^{-}}\bra{\psi^{-}},
\end{equation}
where $e_{mA}$ is the probability of a misalignment error. The same is true on Bob's side as well with misalignment error probability $e_{mB}$. Since the middle station is located halfway between the parties, we will assume that
\begin{equation}
e\coloneqq e_{mA}=e_{mB}.
\end{equation}
\subsubsection{Detectors}
Alice and Bob both have threshold detectors with dark count probability $p_d$. To perform the BB84 measurements Alice and Bob need to use a detector setup consisting of two detectors and an optical device that can distinguish between the eigenstates of the $X$ and $Z$ operators (for instance a polarizing beam splitter). The efficiency of the detectors included in such a setup is denoted by $\eta_d$. The effect of dark counts is taken into account by replacing the quantum state $\rho$ of the photon heading towards the BB84 detection setup with the following state
\begin{equation}\label{darkcount}
\alpha(\eta)\rho+(1-\alpha(\eta))\frac{\mathds{1}}{2},
\end{equation}
where $\eta$ represents the probability that the photon reaches the BB84 detection module and 
\begin{equation}\label{alphaMainText}
\alpha(\eta)=\frac{\eta (1-p_d)}{1-(1-\eta)(1-p_d)^2}.
\end{equation}
\subsubsection{Bell state measurement}
The success probability of the BSM is $p_{BSM}$. We also introduce the BSM ideality parameter $\lambda_{BSM}$, which describes how close is the performed BSM to a perfect BSM. It is modelled by applying a depolarizing channel to the QM states before entering an ideal BSM, therefore the state entering into a perfect BSM instead of $\rho$ is given by
\begin{equation}
\lambda_{BSM}\rho+(1-\lambda_{BSM})\frac{\mathds{1}}{4}.
\end{equation} 
\section{Secret key rate}\label{methods}
The secret key rate of the extended protocol is given by the usual formula
\begin{equation}\label{KeyRateMainText}
R=\frac{Y}{2}\left[1-h(e_X)-f h(e_Z)\right],
\end{equation}
where $Y$ is the yield, which in our case means the expected number of produced raw key bits through steps 1-4 of the protocol per channel use. The function $h$ is the binary entropy function, $e_X$ ($e_Z$) is the quantum bit error rate (QBER) in the $X$ ($Z$)-basis and $f$ is the error correction inefficiency function. We remark that this is not the optimal protocol to extract secret key in this setting and several improvements such as the use of the six-state protocol~\cite{SixState,SixState2}, as used in~\cite{WehnerOptimal}, or noisy preprocessing~\cite{KrausPreProcess,RennerThesis} could be used but this formula is good enough for our purposes, that is, to show the amount of improvement compared to the protocol in~\cite{Luong}. As mentioned previously, the factor $1/2$ is included since two optical modes are used. The channel uses are counted in a similar way as in the original proposal, that is, it is taken to be a product of $m$ and the maximum of the number of rounds that Alice and Bob have to perform in order to obtain at least one raw key bit. 

On the one hand, having $m>1$ makes the probability higher that success is declared in each round for a given distance, which means that once Alice declares success, the QMs on her side will dephase for less time since Bob also have a higher chance to declare success in each round on his side compared to the $m=1$ case, consequently the error rate will decrease. On the other hand, it is possible that in steps 1-4 more than one raw key bits are obtained. Therefore, when deriving the secret key rate we have to take into account the expected number of QM pairs that we can obtain, similarly to how it is done in~\cite{koji}.

The exact derivation of the yield $Y$ and the error rates $e_X$ and $e_Z$ in the case of multiple quantum memories for the secret key rate formula can be found in the Appendix. We remark that, to obtain the secret key rate we made use of the results presented in the original proposal~\cite{Luong} and in~\cite{koji}, which is a QKD protocol similar to our extension, however it can be implemented without using QMs.

\section{Results}\label{Sec:results}
In this chapter, we numerically evaluate the secret key rate formula, obtained in the Appendix and compare it to the PLOB~\cite{PLOB} bound in two cases. In the first case, the photons emitted by the QMs are converted to telecommunication wavelength and in the second case we omit this conversion step, therefore the photons emitted by the QMs enter the fiber at their original wavelength, which means that the corresponding loss in the channel can be very high for these non-converted photons. Our investigations show, that the most crucial parameters of the setup are the dephasing time constant of the QMs, $T_2$, and the overall efficiency $\eta_{total}=\eta_p \eta_c \eta_d$, which is defined as the product of the entanglement preparation efficiency, the coupling efficiency and the detection efficiency. For both cases we fix the values of the following parameters:

\begin{itemize}
\item $T_p=2\,\rm{\mu s}$ (entanglement preparation time)
\item $c=2\cdot10^8\,\rm{m/s}$ (speed of light in the optical fiber) 
\item $p_d=1.8\cdot 10^{-11}$ (dark count probability per detector) 
\item $e=1$ \% (misalignment error)
\item $f=1.16$ (error correction inefficiency)
\item $\lambda_{BSM}=0.98$~\cite{BSMIdeality} (BSM ideality parameter)
\item $p_{BSM}=1$ (BSM success probability)
\end{itemize}

\subsection{With wavelength conversion}\label{Sec:conv}
\begin{figure}[H]
\centering
{\includegraphics[scale=0.62]{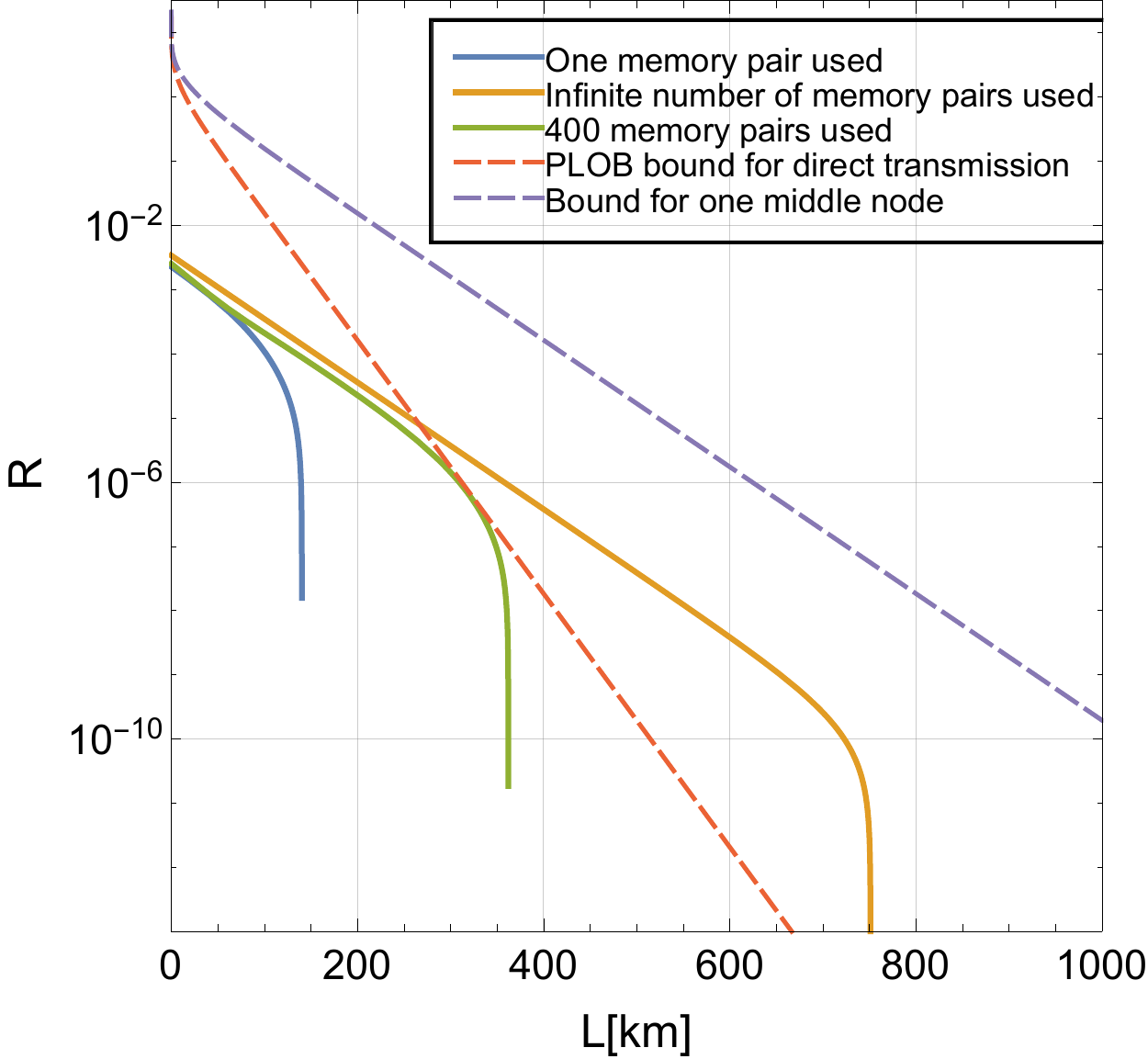}}  
 \caption{Secret key rate versus distance for $m=1$, $m=400$ and $m\rightarrow \infty$ with $\eta_{total}=0.01155$ and $T_2=2\,\rm{s}$ (see below for more details) compared to the PLOB bound and the upper bound for the secret key rate capacity of any protocol with one middle node, placed halfway between the parties~\cite{pirandolaGeneralBound}. The bounds are represented by dashed curves.}
\label{fig:skr}
\end{figure}

First, to have a better sense of the amount of improvement that our extension can offer compared to the original proposal, in Fig.~\ref{fig:skr} we plot the secret key rate of the extended protocol for the special cases of $m=1$, $m=400$ and $m\rightarrow \infty$ for a specific $T_2$ and $\eta_{total}$ value (see below) together with the PLOB bound and a similar bound corresponding to secret key rate capacity of QKD protocols that contain one middle node~\cite{pirandolaGeneralBound}. As mentioned previously, in this case the photons emitted by the QMs are converted to telecommunication wavelength so we set $L_{att}=22\,\rm{km}$ corresponding to low-loss in the channel. Apart from the previously set parameters, we use the following values for the remaining parameters, similarly to~\cite{Luong}:
\begin{itemize}
\item $T_2=2\,\rm{s}$~\cite{T2Ref} (dephasing time constant)
\item $\eta_{p}=0.66$ (entanglement preparation efficiency)
\item $\eta_{c}=0.05 * 0.5$~\cite{CouplingRef} (photon-fiber coupling efficiency * wavelength conversion efficiency)
\item $\eta_{d}=0.7$ (detection efficiency)
\item These values result in $\eta_{total}=\eta_{p}\eta_{c}\eta_{d}=0.0115$.
\end{itemize} 

We can already see in Fig.~\ref{fig:skr} that increasing the number of applied memory modules ($m$) can make overcoming the PLOB bound possible, which was not feasible in the case of $m=1$ for the above listed $T_2$ and $\eta_{total}$ parameter values. Indeed, the beating happens around $m=400$. Of course, we do not stand a chance against the fundamental secret key capacity bound for QKD protocols containing one middle node since in our scheme the middle node comes with many imperfections. We note, that the general ($m\geq 1$) secret key rate formula derived in the Appendix taken in the special case of $m=1$ gives the same curve as in~\cite{Luong}.
 
\begin{figure}[H]
\centering
{\includegraphics[scale=0.4]{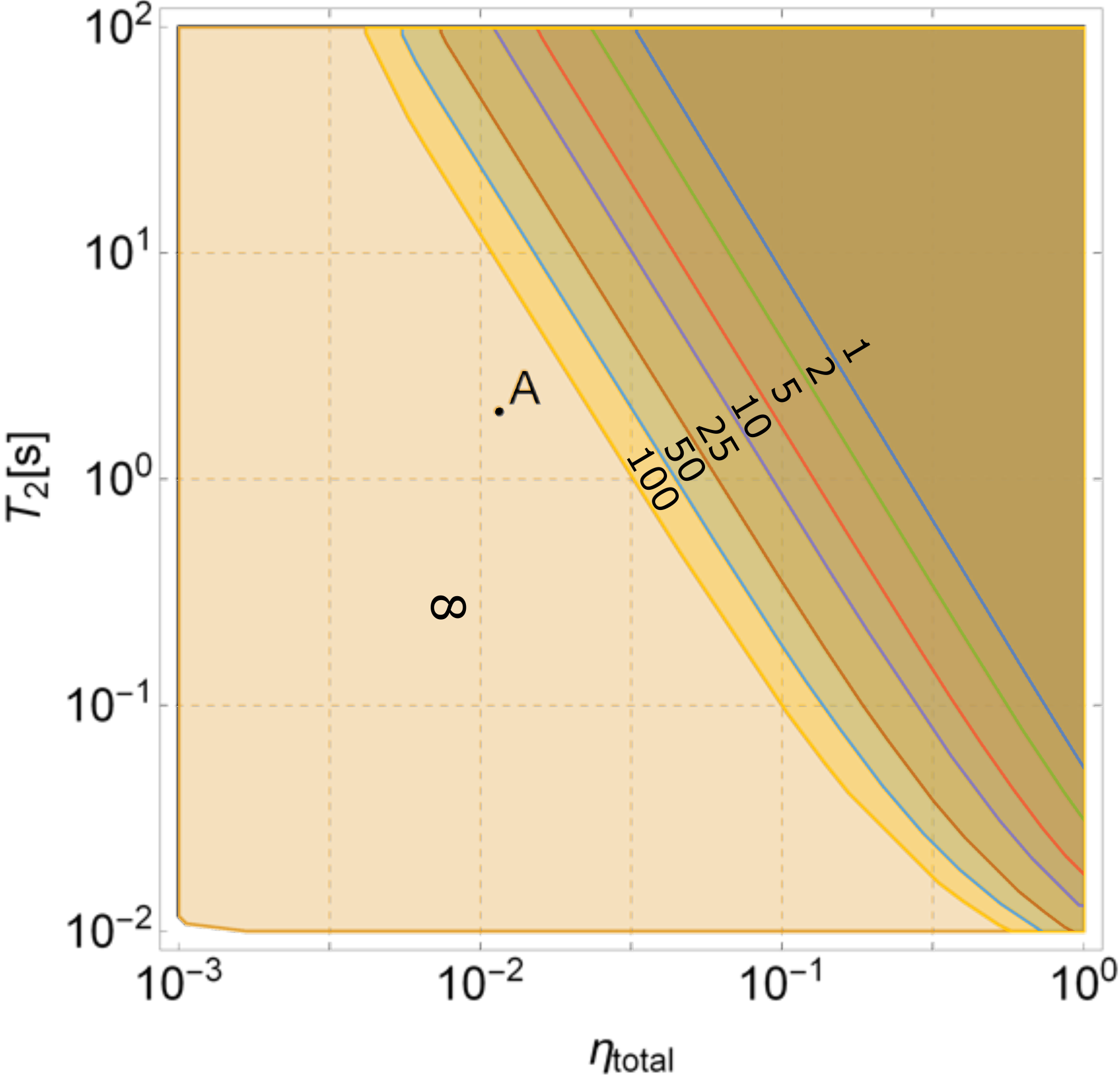}} 
 \caption{Regions in the $\eta_{total}-T_2$ memory parameter space where the extended protocol beats the PLOB bound, given that Alice and Bob have $m$ pairs of memories between them. Point A has coordinates $\eta_{total}=0.0115$, $T_2=2\,\rm{s}$.}
\label{fig:plobinf}
\end{figure}

Now, we take a grid of points in the QM parameter space $\eta_{total}$-$T_2$ and for each point ($\eta_{total}$, $T_2$), given the number of memory modules $m$, we determine if the PLOB bound is overcome by the extended protocol. The result of the numerical simulations appears in Fig.~\ref{fig:plobinf}. We note that the secret key rate curves in Fig.~\ref{fig:skr} correspond to the point A ($\eta_{total}=0.0115$, $T_2=2\,\rm{s}$) in Fig.~\ref{fig:plobinf}, which leads to a crossover for $m\approx 400$. 
We have already seen in Fig.~\ref{fig:skr} that adding more memory modules can relax the required quality of the QMs (in terms of $T_2$ and $\eta_{total}$) for our protocol to be able to beat the PLOB bound. In Fig.~\ref{fig:plobinf}, we have confirmed and quantified this effect in the whole parameter space $\eta_{total}$-$T_2$. Therefore, given the parameters of the QMs at one's disposition one can check the required number of memory stations to beat the PLOB bound. Similarly, if we know how many QM modules can be handled in a parallel manner in our system, then we can read off from Fig.~\ref{fig:plobinf} how to improve the QM parameters to beat the PLOB bound. By looking at the slope of the lines bordering the regions in Fig.~\ref{fig:plobinf} we observe that we need a smaller relative improvement in the parameter $\eta_{total}$ than in $T_2$ to step into regions corresponding to smaller values of $m$. However, it can also be seen that the improvement is moderate, in order to have a significant relaxation on the required parameters $T_2$ and $\eta_{total}$ the number of memory modules has to be very high, which is probably more challenging to manage experimentally than achieving better $T_2$ and $\eta_{total}$ values for the QMs in the first place.

\subsection{Without wavelength conversion}\label{Sec:withoutconv}
\begin{figure}[H]
\centering
\subfigure[]{\includegraphics[scale=0.35]{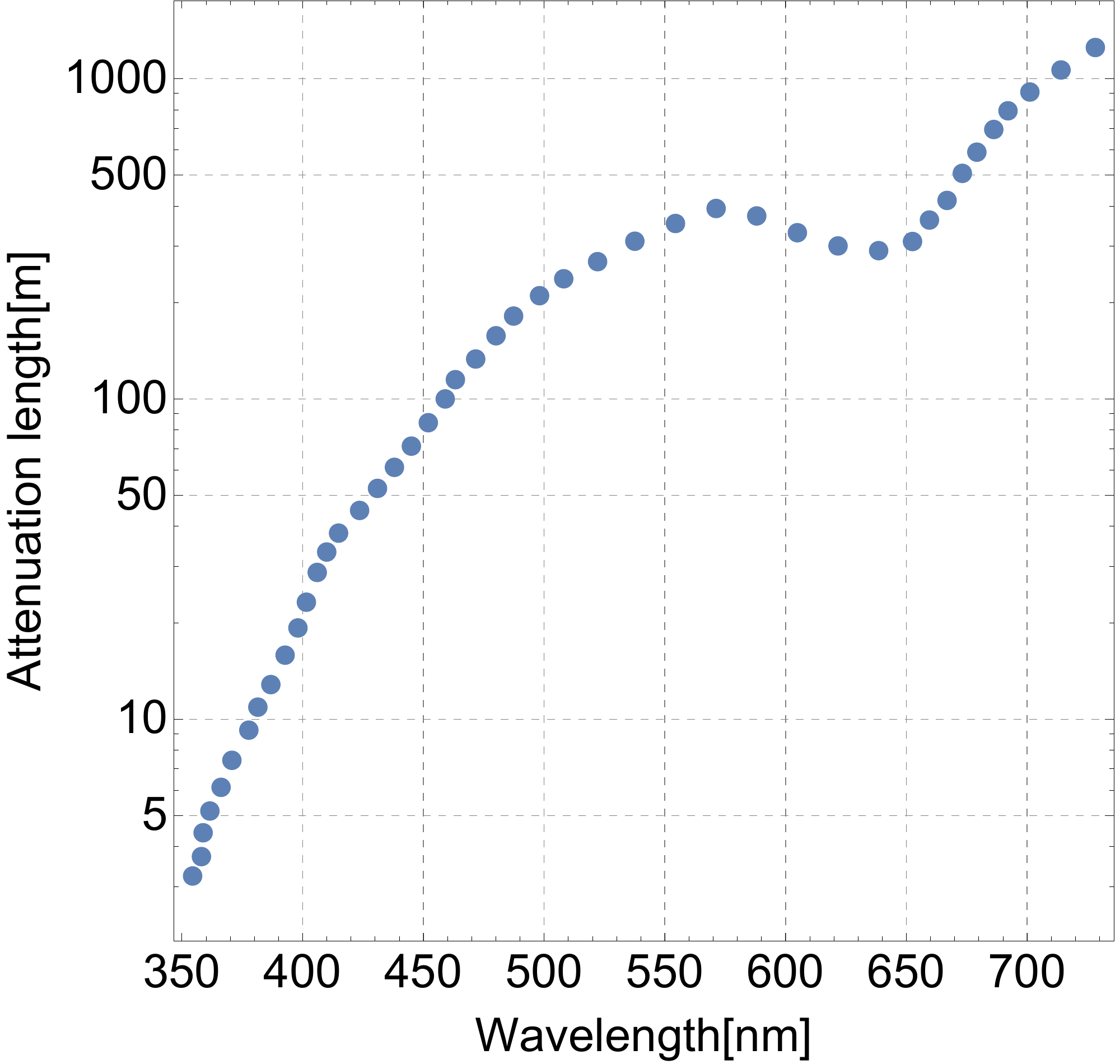}}
\label{fig:HL1a}
\subfigure[]{\includegraphics[scale=0.35]{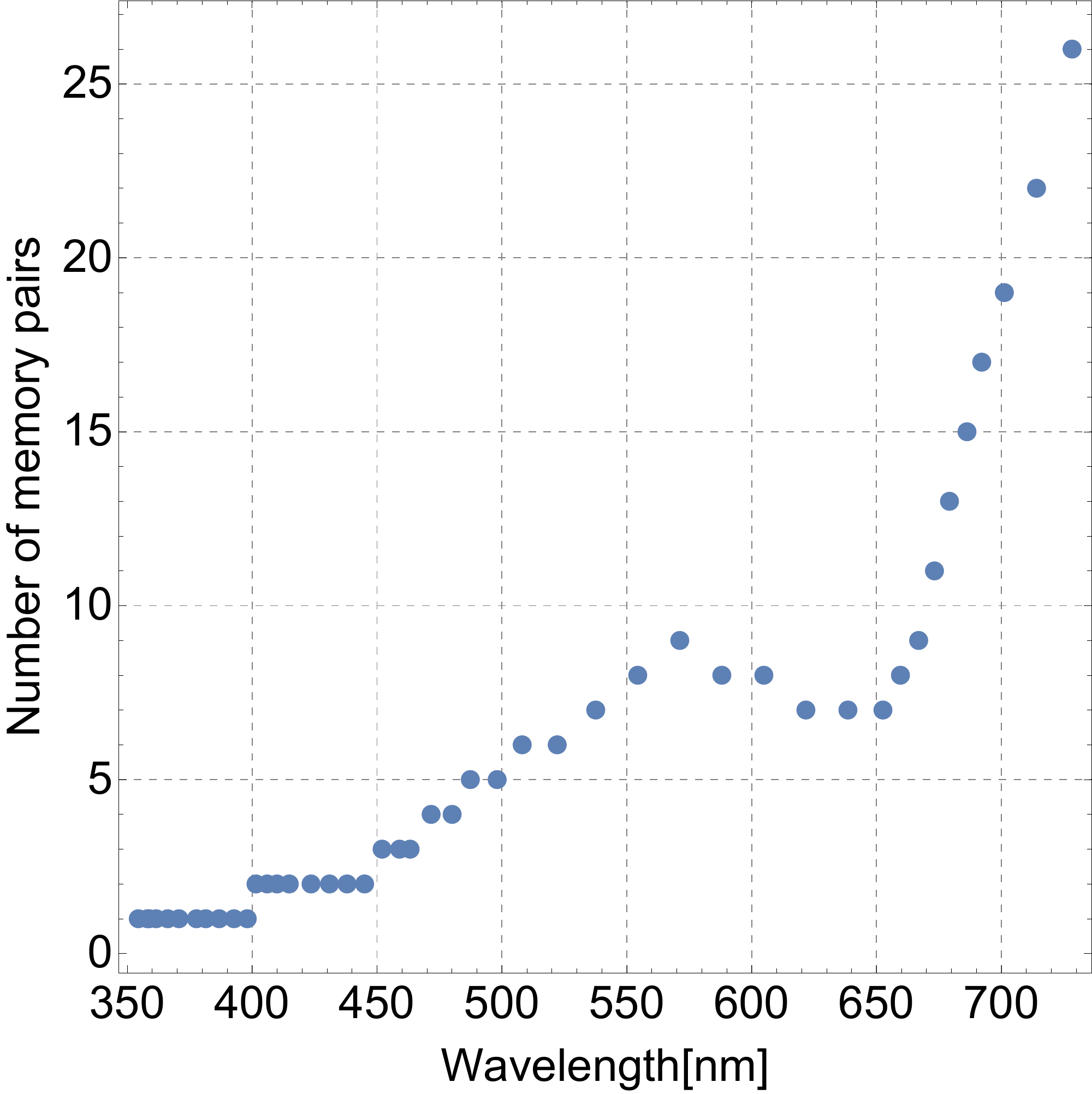}}
\label{fig:HL1b} 
  \caption{(a) Attenuation length versus wavelength for the applied fiber~\cite{FiberExample}. (b) The minimum number of the necessary memory pairs for beating the PLOB bound, as a function of the wavelength of the emitted photons, assuming that the optical fiber~\cite{FiberExample} is employed. We set $T_2=2\,\rm{s}$ and $\eta_{total}=0.01155$.}
  \label{fig:HighLoss}
\end{figure}

In this section, we consider the high-loss channel scenario (\textit{i.e.}, small $L_{att}$). This makes surpassing the PLOB bound easier, since the emitted photons will not travel so far, therefore the loaded QMs will dephase for less time. This means that high-loss already relaxes the conditions on the parameters $T_2$ and $\eta_{total}$, so overcoming can be made possible with adding less memory modules than in the low-loss scenario. For the simulations, we take the attenuation length over wavelength data (see Fig.~\ref{fig:HighLoss}a) from a commercially available optical fiber~\cite{FiberExample} and we assume that the wavelength of the photons emitted by the QMs falls in this range, moreover, the photons are directed into the optical fiber without wavelength-conversion. In this setting, we obtain numerically the minimum number of required memory modules for beating the PLOB bound as a function of the emitted wavelengths. In the simulations, to have the same $\eta_{total}$ value as in Fig.~\ref{fig:skr} we set $\eta_{c}=0.025 * 1$, with the conversion efficiency being 1 since the conversion step is omitted. The other parameters have the same values as in the previous sections. The results can be seen in Fig.~\ref{fig:HighLoss}.
In Fig.~\ref{fig:HighLoss}b, we see, that in the high-loss scenario, adding just a couple more modules can help beating the PLOB bound which was not possible before with $m=1$. Note, however, that in this regime using the optical fiber~\cite{FiberExample} the distance between the parties is comparatively short (in the order of hundreds of meters). Nevertheless, implementing a QKD protocol that can beat the PLOB bound would be a relevant result no matter how small is the distance between the parties.
\section{Conclusion}\label{Sec:Conclusion}
In this paper, we have evaluated the performance of the protocol proposed in~\cite{Luong} when extended with multiple QM pairs, operated in a parallel way. We have determined how the number of applied QM pairs affects the required QM parameters (e.g., the dephasing time constant) for beating the PLOB bound. We have found and described quantitatively that with this extension one can beat the PLOB bound with lower quality QMs. The advantage compared to the original protocol~\cite{Luong} comes from the fact that there are more QMs available in a round, therefore it is more probable that the parties will load at least one of the QMs on their side, so the pairings and the BSMs between the loaded QMs can start sooner, which means that the QMs will dephase for less time. 

However, to have a significant relaxation on the QM parameters one needs to use many QM pairs. At this point we also note that implementing this protocol could be challenging due to the multiple memory modules and the matching mechanism needed, however, we believe that it should still be easier than realizing a full-scale quantum repeater. We have also explored the high-loss limit, in which, adding just a few more QMs can actually make the crossover possible since the photons do not travel so far, therefore there is less time for dephasing in the QMs in the first place. 

\section{Acknowledgment}\label{Sec:Ack}

We thank the European Union's Horizon 2020 research and innovation program under the Marie Sklodowska-Curie grant agreement No 675662 (project QCALL) and the ARL CDQI program for financial support.

\appendix*\label{Appendix}
\section{Derivation of the secret key rate}
In this appendix, first, we recall the secret key rate formula from~\cite{Luong} for the original ($m=1$) case, assuming that the memory station is located halfway between Alice and Bob, who are separated at a distance $L$. Then, we describe how to modify the quantities therein for obtaining the secret key rate formula for the $m>1$ case.
\subsection{Results for the $m=1$ case}
The secret key rate for the $m=1$ case~\cite{Luong} is lower bounded~\cite{KeyrateShor,KeyrateNorbert} by the quantity 
\begin{equation}\label{OriginalKeyRate}
R=\frac{Y}{2}\left[1-h(e_X)-f h(e_Z)\right],
\end{equation}
where $Y$ is the yield, which represents the probability per channel use that the parties obtain a raw key bit. $h$ is the binary entropy function, $e_X$ ($e_Z$) is the quantum bit error rate (QBER) in the $X$ ($Z$)-basis and $f$ is the error correction inefficiency function. 

The probability that a photon (emitted by a memory module) is detected is given by
\begin{equation}\label{etaOriginal}
\eta=\eta_{total} e^{-\frac{L}{2L_{att}}}.
\end{equation} 
Taking into account the dark counts, the probability that a detector clicks can be written as
\begin{equation}\label{etaPrimeOriginal}
\eta'=1-(1-\eta)(1-p_d)^2,
\end{equation}
where $p_d$ is the dark count probability of a detector. Let $N_A$ ($N_B$) denote the number of trials until Alice's (Bob's) detector clicks. Note, that $N_A$ ($N_B$) is a geometrically distributed random variable with success probability parameter $\eta'$.
The yield in the original case was shown to be~\cite{Luong,panayi} 
\begin{equation}\label{YieldOriginal}
Y=\frac{p_{BSM}}{\mathbb{E}[\max(N_A,N_B)]}=p_{BSM}\left(\frac{2}{\eta'}-\frac{1}{2\eta'-(\eta')^2} \right)^{-1},
\end{equation} 
where $\mathbb{E}$ is the expectation value operator and $p_{BSM}$ is the success probability of the BSM between the memories. The quantity $\mathbb{E}[\max(N_A,N_B)]$ in the denominator of Eq.~\eqref{YieldOriginal} represents the average number of channel uses.

Now, we recall the QBERs for the $m=1$ case. In~\cite{Luong}, it is shown, that 
\begin{align}\label{ex}
e_X&=\lambda_{BSM} \alpha^2(\eta)\left[\epsilon_{mis}(1-\epsilon_{dp})+\epsilon_{dp}(1-\epsilon_{mis})\right]\nonumber\\&+\frac{1-\lambda_{BSM} \alpha^2(\eta)}{2},
\end{align}
where $\lambda_{BSM}$ is the BSM ideality parameter, the function $\alpha$ is given by Eq.~\eqref{alphaMainText},
\begin{equation}
\epsilon_{mis}=2e(1-e),
\end{equation}
where $e$ is the misalignment error in one channel. Moreover
\begin{equation}\label{dephasingError}
\epsilon_{dp}=\mathbb{E}[\lambda_{dp}(t_A)[1-\lambda_{dp}(t_B)]+\lambda_{dp}(t_B)[1-\lambda_{dp}(t_A)]].
\end{equation}
The quantities $\epsilon_{mis}$ and $\epsilon_{dp}$ are regarded as the total misalignment and dephasing errors in the system. The function $\lambda_{dp}$ is given by Eq.~\eqref{dephasingLambda} and, as mentioned previously, describes the dephasing of the state stored in the quantum memory over time. The quantity $t_A$ ($t_B$) is the time that the loaded quantum memory on Alice's (Bob's) side is left to dephase for. The time that elapses between the rounds is 
\begin{equation}
\tau=T_p+\frac L c,
\end{equation} 
where $T_p$ is the preparation time for a QM and $c$ denotes the speed of light in the optical fibers used. The term $L/c$ comes from the fact that the signals need to reach Alice (or Bob) and they have to report back to the memory station whether they obtained a successful detection or not. With this quantity we have that
\begin{equation}\label{taOriginal}
t_A=N_B\tau+\frac L c
\end{equation}
and
\begin{equation}\label{tbOriginal}
t_B=\frac L c.
\end{equation}
After Alice gets a successful detection on her side, the memories have already been dephasing for $L/(2c)$ time, on the top of this, reporting the success to the middle station takes at least an extra $L/(2c)$ time. Moreover, Alice has to wait until Bob also gets a successful detection so that they can perform the BSM between the memories, this is accounted for in the term $N_B\tau$. Bob does not have to wait for anybody else, so $t_B$ only contains the term $L/c$. For deriving the quantity $\epsilon_{dp}$, one has to calculate the following expectation value $\mathbb{E}[ e^{-t_A/T_2}]$, which has been derived in~\cite{Luong}
\begin{equation}\label{expectedValueOriginal}
\mathbb{E}[ e^{-t_A/T_2}]=\frac{\eta'\exp\left( -\frac{L}{cT_2}\right)}{\exp(\tau/T_2)+\eta'-1}.
\end{equation}
With this expectation value and due to the linearity of the expectation value operator $\epsilon_{dp}$ can be derived easily, with which $e_{X}$ can be obtained.

Moreover, $e_Z$ is given by
\begin{equation}\label{ezOriginal}
e_Z=\lambda_{BSM}\alpha^2(\eta)\epsilon_{mis}+\frac{1-\lambda_{BSM} \alpha^2(\eta)}{2}.
\end{equation}

\subsection{Results for the $m>1$ case}

Here, we explain how to obtain the secret key rate formula for the $m>1$ case, using the previously presented results from~\cite{Luong}. 

The extension of the memory station to contain $m>1$ memory modules means that in step 2 of the protocol, Bob has to perform less rounds until success, since he has a higher chance to detect at least one photon out of the $m>1$ signals than in the original $m=1$ case for a given $L$ distance. Therefore, Alice (who already has at least one loaded memory on her side) has to wait less for Bob, which means that loaded memories on Alice's side will dephase for less time on average, meaning that the error rate will decrease. Thus, all the quantities that include the expectation values of the number of performed rounds has to be changed for the case of $m>1$. Moreover, with this extension, it is possible, that more pairs are obtained in a round, however, at the same time the number of channel uses also increase. Apart from these modifications each pairing is basically the same process as it was for $m=1$, even if in one round more pairings can happen. This means that the structure of the formula given by Eq.~\eqref{OriginalKeyRate} stays valid even for the $m>1$ case, however, one has to apply the modified quantities, that will be given in the remainder of this section.

From Eq.~\eqref{YieldOriginal} and Eq.~\eqref{ex} it is clear, that in Eq.~\eqref{OriginalKeyRate}, $Y$ and $e_X$ have to be modified since these are the quantities that contain the expectation values of waiting times and number of channel uses. Eq.~\eqref{ezOriginal} shows, that $e_Z$ does not depend neither on the expectation value of waiting times nor on the expected number of channel uses, therefore it does not have to be modified. 

Let $p_s$ denote the probability that at least one (out of $m$) of Alice's (or Bob's) detectors clicks, meaning that she (or he) declares success in a round. It is easy to see that 
\begin{equation}\label{generalEta}
p_s=1-(1-\eta')^m.
\end{equation}
We remark that, as expected, $m=1$ implies that $p_s=\eta'$. 
 
Let us keep the notation $N_A$ ($N_B$) to denote the number of trials until Alice (Bob) declares success. In the $m>1$ case, success means that Alice's (Bob's) detectors detect at least one out of the $m$ photons emitted by the memory modules. Therefore, $N_A$ ($N_B$) is still a geometrically distributed random variable, however, with a success probability parameter $p_s$ for the $m>1$ case (which was $\eta'$ for the $m=1$ case). This means that whenever an expectation value of a quantity containing $N_A$ and $N_B$ was calculated for the $m=1$ case, the corresponding expectation value for the $m>1$ case can be obtained from the single memory ($m=1$) case by replacing $\eta'$ with $p_s$ in its expression. 

Let $Y_k$ denote the probability that the parties obtain $k$ sifted bits after the pairings. The yield in the extended case can be written as 
\begin{equation}\label{Yield}
Y=\frac{\sum_{k=1}^{m} k\, Y_k }{m\,\mathbb{E}[\max(N_A,N_B)]},
\end{equation} 
where, similarly to Eq.~\eqref{YieldOriginal} the denominator $m\,\mathbb{E}[\max(N_A,N_B)]$ represents the expected number of channel uses to obtain at least one raw key bit, here $m$ appears since in each round the memory modules emit $m$ signals corresponding to $m$ channel uses in each round. We note, that since we use polarization encoding, we need two modes per qubit signal but this is already accounted for in the secret key rate formula given by Eq.~\eqref{OriginalKeyRate} as a factor 1/2. The quantity $\mathbb{E}[\max(N_A,N_B)]$ is given in Eq.~\eqref{YieldOriginal} for the $m=1$ case. Due to our previous remark, for the extended case it can be written as (compare with Eq.~\eqref{YieldOriginal})
\begin{equation}
m\,\mathbb{E}[\max(N_A,N_B)]=m\left(\frac{2}{p_s}-\frac{1}{2p_s-p_s^2} \right)
\end{equation}

Now, we derive $Y_k$. The derivation is basically the same as in~\cite{koji}. Let us introduce
\begin{equation}\label{binom}
B_i^m(p)=\binom{m}{i} p^i(1-p)^{m-i},
\end{equation} which is the probability mass function of the binomial distribution with parameter $0<p<1$. Let $k_A$ ($k_B$) denote the number of successful detections on Alice's (Bob's) side in a successful round, remember that a round is successful when there is at least one successful detection. The probability that $\min(k_A,k_B)=l$, which is exactly the number of pairs that we can obtain, is given by
\begin{equation}\label{MinEqualsL}
p_{l=\min(k_A,k_B)}=2B_l^m(\eta')\sum_{x=l}^{m}B_x^m(\eta')-(B_l^m(\eta'))^2.
\end{equation}  
This is so since either we have $l$ detections at Alice's side and $l,\, l+1,\, \dots,\, m$ detections at Bob's side or the other way around (this introduces the factor 2 in Eq.~\eqref{MinEqualsL}), but the case when they both have $l$ clicks is counted twice, therefore this has to be subtracted.

The probability to make $k$ pairs can be written as
\begin{equation}\label{pk}
p_{k\rm{bits}}=\sum_{y=k}^{m} p_{y=\min(k_A,k_B)}B_k^y(p_{BSM}).
\end{equation}
This expression holds since if the minimum of the successful detections at Alice's and Bob's side is $y\geq k$ then, out of the possible $y$ pairs we obtain $k$ pairs with probability $B_k^y(p_{BSM})$. With this we have that
\begin{equation}
Y_k=\sum_{a=1}^{\infty}\sum_{b=1}^{\infty}(1-p_s)^{a-1}(1-p_s)^{b-1} p_{k\rm{bits}}=\frac{p_{k\rm{bits}}}{p_s^2},
\end{equation}
where we summed over the possible number of rounds that the parties have to perform until they both have at least one successful detection. Thus we have that
\begin{equation}\label{Y}
Y=\frac{\frac{1}{p_s^2}\sum_{k=1}^{m} k\,p_{k\rm{bits}} }{m\left(\frac{2}{p_s}-\frac{1}{2p_s-p_s^2} \right)},
\end{equation}
which can now be calculated with Eq.~\eqref{MinEqualsL} and Eq.~\eqref{pk}. We note, that in~\cite{koji}, $\sum_{k=1}^{m} k\,p_{k\rm{bits}}$ is written in the following different form:
\begin{equation}\label{kojiSimpler}
\sum_{k=1}^{m} k\,p_{k\rm{bits}}=mp_{BSM}[\eta'-g_m(\eta')],
\end{equation}
where
\begin{align}
g_m(\eta')&=\eta'(1-\eta')\biggr[ \sum_{i=0}^{m-1}\left[B_i^{m-1}(\eta')\right]^2\nonumber\\
&+\sum_{i=1}^{m-1}B_i^{m-1}(\eta')B_{i-1}^{m-1}(\eta')\biggr]
\end{align}
from which it is easier to extract the expression for $Y$ in the case of $m\rightarrow\infty$ since $g_m\rightarrow 0$ as $m\rightarrow \infty$ as it is shown in~\cite{koji}.

Now, we adapt $e_X$, given by Eq.~\eqref{ex}, for the $m>1$ case. In doing so, we only have to modify the total dephasing error $\epsilon_{dp}$, given by Eq.~\eqref{dephasingError}, since this is the quantity that contains the expectation value of the number of performed rounds. For this, we have to calculate expectation values of the form $\mathbb{E}[ e^{-t_A/T_2}]$, where $t_A$ is the amount of time that Alice's loaded QMs are left to dephase for and is still given by Eq.~\eqref{taOriginal}. This expectation value is given by Eq.~\eqref{expectedValueOriginal} for the $m=1$ case. As we have pointed out before, to generalize the expectation value from the $m=1$ case to the $m>1$ case, we simply have to replace $\eta'$ with $p_s$ in Eq.~\eqref{expectedValueOriginal}:
\begin{equation}\label{expectedValue}
\mathbb{E}[ e^{-t_A/T_2}]=\frac{p_s\exp\left( -\frac{L}{cT_2}\right)}{\exp(\tau/T_2)+p_s-1}.
\end{equation}
Using this, we can express $\epsilon_{dp}$ for the $m>1$ case as
\begin{align}\label{dephasingFinal}
\epsilon_{dp}&=\left(1-2\lambda_{dp}\left(\frac L c\right)\right)\left(\frac 12 - \frac 1 2 \frac{p_s\exp\left(-\frac{L}{c T_2}\right)}{\exp\left(\frac{\tau}{T_2}\right)+p_s-1}\right)\nonumber\\&+\lambda_{dp}\left(\frac L c\right).
\end{align}
As argued previously, $e_Z$ stays the same for the $m>1$ case and is given by Eq.~\eqref{ezOriginal}.

Plugging Eq.~\eqref{Y}, Eq.~\eqref{ex} with Eq.~\eqref{dephasingFinal} and Eq.~\eqref{ezOriginal} into Eq.~\eqref{OriginalKeyRate} we obtain the secret key rate formula of the protocol for $m>1$.


\begin{thebibliography}{99}
\bibitem{GisinReview} N. Gisin, G. Ribordy, W. Tittel and H. Zbinden: Quantum cryptography, Reviews of Modern Physics 74, 145 (2002).
\bibitem{NorbertReview} V. Scarani, H. B.-Pasquinucci, N. J. Cerf, M. Du\v{s}ek, N. L\"utkenhaus and M. Peev: The security of practical quantum key distribution, Reviews of Modern Physics 81, 1301 (2009).
\bibitem{MarcosReview} H.-K. Lo, M. Curty and K. Tamaki: Secure quantum key distribution, Nature Photonics 8, 595-604 (2014).
\bibitem{Luong} D. Luong, L. Jiang, J. Kim, and N. L\"utkenhaus: Overcoming lossy channel bounds using a single quantum repeater node, Applied Physics B 122, 96 (2016).
\bibitem{Pir0} S. Pirandola, R. Garc\'ia-Patr\'on, S. L. Braunstein and S. Lloyd: Direct and Reverse Secret-Key Capacities of a Quantum Channel, Physical Review Letters 102, 050503 (2009).
\bibitem{TGW} M. Takeoka, S. Guha, and M. M. Wilde: Fundamental rate-loss tradeoff for optical quantum key distribution, Nature Communications 5, 5235 (2014).
\bibitem{AnotherPaperOnBounds} M. M. Wilde, M. Tomamichel and M. Berta: Converse Bounds for Private Communication Over Quantum Channels IEEE Transactions on Information Theory 63, 1792-1817 (2017).
\bibitem{PLOB} S. Pirandola, R. Laurenza, C. Ottaviani, and L. Banchi: Fundamental limits of repeaterless quantum communications, Nature Communications 8, 15043 (2017).
\bibitem{Pir1} S. Pirandola \textit{et} \textit{al}.: Theory of channel simulation and bounds for private communication, Quantum Science and
Technology 3, 040101 (2018).

\bibitem{mdi} H.-K. Lo, M. Curty, and B. Qi: Measurement-Device-Independent Quantum Key Distribution, Physical Review Letters 108, 130503, (2012).


\bibitem{LucamariniTfQkd} M. Lucamarini, Z. L. Yuan, J. F. Dynes and A. J. Shields: Overcoming the rate–distance limit of quantum key distribution without quantum repeaters, Nature 557, 400–403 (2018).
\bibitem{CuiTF} C. Cui \textit{et} \textit{al}.: Twin-field quantum key distribution without phase post-selection, preprint arXiv:1807.02334, (2018).
\bibitem{JieTF} J. Lin and N. L\"utkenhaus: Simple security analysis of phase-matching measurement-device-independent quantum key distribution, Physical Review A 98, 042332 (2018).
\bibitem{MarcosTf} M. Curty, K. Azuma, H.-K. Lo: Simple security proof of twin-field type quantum key distribution protocol, npj Quantum Information 5, 64 (2019).

\bibitem{ProofOfPrincipleTF} X. Zhong, J. Hu, M. Curty, L. Qian and H.-K. Lo: Proof-of-principle experimental demonstration of twin-field type quantum key distribution, preprint arXiv:1902.10209v1 (2019).
\bibitem{ToshibaExpTF} M. Minder \textit{et} \textit{al.}: Experimental quantum key distribution beyond the repeaterless secret key capacity, Nature Photonics 13, 334-338 (2019) 
\bibitem{PanTF} Y. Liu \textit{et} \textit{al.}: Experimental Twin-Field Quantum Key Distribution Through Sending-or-Not-Sending, preprint arXiv:1902.06268v1 (2019).
\bibitem{WangTFExp} S. Wang \textit{et} \textit{al.}: Beating the Fundamental Rate-Distance Limit in a Proof-of-Principle
Quantum Key Distribution System, Physical Review X 9, 021046 (2019).

\bibitem{BriegelRepeater} H.-J. Briegel, W. D\"ur, J. I. Cirac and P. Zoller: Quantum repeaters: the role of imperfect local operations in quantum communication, Physical Review Letters 81, 5932 (1998).
\bibitem{DuanRepeater} L.-M. Duan, M. D. Lukin, J. I. Cirac and P. Zoller: Long-distance quantum communication with atomic ensembles and linear optics, Nature 414, 413–418 (2001).
\bibitem{HybridRepeater} P. van Loock, T. D. Ladd, K. Sanaka, F. Yamaguchi, K. Nemoto, W. J. Munro and Y. Yamamoto: Hybrid Quantum Repeater Using Bright Coherent Light, Physical Review Letters 96, 240501 (2006).
\bibitem{KokRepeater} P. Kok, C. P. Williams and J. P. Dowling: Construction of a quantum repeater with linear optics, Physical Review A 68, 022301 (2003).

\bibitem{koji} K. Azuma, K. Tamaki and W. J. Munro: All-photonic intercity quantum key distribution, Nature Communications 6:10171 (2015).

\bibitem{dephasingproblem} B. Zhao \textit{et} \textit{al}.: A millisecond quantum memory for scalable quantum networks, Nature Physics 5, 95-99 (2009).

\bibitem{abruzzo} S. Abruzzo, H. Kampermann and D. Bru{\ss}: Measurement-device-independent quantum key distribution with quantum memories, Physical Review A 89, 012301 (2014).
\bibitem{panayi} C. Panayi, M. Razavi, X. Ma and N. L\"utkenhaus: Memory-assisted measurement-device-independent quantum key distribution, New Journal of Physics 16, 043005 (2014).
\bibitem{ChristophSimon} S. Guha, H. Krovi, C. A. Fuchs, Z. Dutton, J. A. Slater, C. Simon and W. Tittel: Rate-loss analysis of an efficient quantum repeater architecture, Physical Review A 92, 022357 (2015).
\bibitem{razavi} N. L. Piparo, M. Razavi and W. J. Munro: Measurement-device-independent quantum key distribution with nitrogen vacancy centers in diamond, Physical Review A 95, 022338 (2017). 
\bibitem{razaviMultiple} N. L. Piparo, N. Sinclair and M. Razavi: Memory-assisted quantum key distribution resilient against multiple-excitation effects, Quantum Science and Technology 3, 1 (2017).
\bibitem{dephasing} M. Razavi, M. Piani, and N. L\"utkenhaus: Quantum repeaters with imperfect memories: Cost and scalability,
Physical Review A 80, 032301 (2009).

\bibitem{SixState} H. Bechmann-Pasquinucci and N. Gisin: Incoherent and coherent eavesdropping in the six-state protocol of quantum cryptography, Physical Review A 59, 4238-4248 (1999).
\bibitem{SixState2} D. Bru{\ss} and and C. Macchiavello: Optimal Eavesdropping in Cryptography with Three-Dimensional Quantum States, Physical Review Letters 88, 127901 (2002).
\bibitem{WehnerOptimal} F. Rozpedek \textit{et} \textit{al}.: Near-term quantum-repeater experiments with nitrogen-vacancy centers: Overcoming the limitations of direct transmission, Physical Review A 99, 052330 (2019).
\bibitem{KrausPreProcess} B. Kraus, N. Gisin and R. Renner: Lower and Upper Bounds on the Secret-Key Rate for Quantum Key Distribution Protocols Using One-Way Classical Communication, Physical Review Letters 95, 080501 (2005).
\bibitem{RennerThesis} R. Renner: Security of Quantum Key Distribution, PhD thesis  arXiv:quant-ph/0512258v2 (2006). 

\bibitem{FiberExample} https://www.thorlabs.com/
thorproduct.cfm?partnumber=FG050LGA

\bibitem{effqkd}H.-K. Lo, H.F. Chau and M. Ardehali: Efficient Quantum Key Distribution Scheme and a Proof of Its Unconditional Security, Journal of Cryptology 18, 133-165 (2005).
\bibitem{poster} E. M.-Scott: Quantum buffering for time-multiplexed multi-photon entanglement, poster presented at CEWQO2019.
\bibitem{BSMIdeality} J. Benhelm, G. Kirchmair, C. F. Roos and R. Blatt: Towards fault-tolerant quantum computing with trapped ions, Nature Physics 4, 463–466 (2008).
\bibitem{T2Ref} S. Olmschenk, K. C. Younge, D. L. Moehring, D. N. Matsukevich, P. Maunz and C. Monroe: Manipulation and detection of a trapped Yb+ hyperfine qubit, Physical Review A 76, 052314 (2007).
\bibitem{CouplingRef} E. W. Streed, B. G. Norton, A. Jechow, T. J. Weinhold and D. Kielpinski: Imaging of Trapped Ions with a Microfabricated Optic for Quantum Information Processing, Physical Review Letters 106, 010502 (2011).
\bibitem{pirandolaGeneralBound} S. Pirandola: End-to-end capacities of a quantum communication network, Communications Physics 2, 51 (2019). 
\bibitem{KeyrateShor} P. W. Shor and J. Preskill: Simple proof of security of the BB84 quantum key distribution protocol, Physical Review Letters 85, 441–444 (2000).
\bibitem{KeyrateNorbert} D. Gottesman, H.-K. Lo, N. L\"utkenhaus and J. Preskill: Security of quantum key distribution with imperfect devices, Quantum Information and Computation 4,	325-360 (2004). 

\end{thebibliography}
\end{document}